\documentstyle[12pt]{article}

\input epsf

\begin{document}

\begin{titlepage}

\hspace{10.3cm}{\small{CALT-68-1885}}

\hspace{10.3cm}{\small{DOE RESEARCH AND}}

\hspace{10.3cm}{\small{DEVELOPMENT REPORT}}

\vspace{2cm}

\begin{center}

{\bf DILATON BLACK HOLES WITH ELECTRIC CHARGE}
\footnote[1]{Work supported in part by the U.S. Dept.
of Energy under Grant No. DE-FG03-92-ER-40701.}

\vspace{1.5cm}

{Malik~Rakhmanov}

\vspace{1cm}

{\it California Institute of Technology\\
Pasadena, CA 91125}

\vspace{2cm}

\end{center}

\abstract{
Static spherically symmetric solutions of the Einstein-Maxwell
gravity with the dilaton field are described. The solutions
correspond to black holes and are generalizations of the
previously known dilaton black hole solution. In addition to
mass and electric charge these solutions are labeled by a new
parameter, the dilaton charge of the black hole. Different
effects of the dilaton charge on the geometry of space-time
of such black holes are studied. It is shown that in most
cases the scalar curvature is divergent at the horizons.
Another feature of the dilaton black hole is that there is a
finite interval of values of electric charge for which no
black hole can exist.
}

\end{titlepage}

\section{Introduction}

The low-energy limit of string theory includes a scalar
dilaton field, which is massless in all finite orders
of perturbation theory \cite{Green}. However, in order
not to conflict with classical tests of the tensor
character of gravity, the physical dilaton should have
mass.  This mass is assumed to arise due to nonperturbative
features of the quantum theory. At the classical level,
and at distance scales small compared to the dilaton
Compton wavelength, we can neglect the mass and study the
effect of the dilaton on low-energy physics. In particular,
the dilaton modifies Maxwell's equations and it affects
the geometry of the space-time \cite{Brans}. For example,
solutions that correspond to electrically charged black holes
are modified by the presence of the dilaton. Such solutions
were studied in \cite{Boulware}-\cite{GHS}, where it was shown
that the dilaton changes the causal structure of the black
hole and leads to curvature singularities at finite radii.
The black hole solution obtained in \cite{GHS} was also
extensively studied in connection with extremal dilaton
black holes. It was argued that such a black hole behaves
like an elementary particle in the sense that its excitation
spectrum has an energy gap \cite{Preskill}-\cite{Holzhey}.

Although the dilaton field naturally arises in string theory
its existence from the point of view of general relativity is
quite problematic. A generic scalar field can violate the
equivalence principle; see the discussion in \cite{Brans}.
On the other hand, in black hole physics inclusion of a
scalar field leads
to appearance of a ``baryon number" associated with the
field. In the case of the dilaton field this is the dilaton
charge. It is generally believed that no parameters other
than mass, electric charge and angular momentum can be
associated with a black hole (see \cite{Bekenstein}).
This conjecture essentially rules out the existence of the
scalar field in the exterior of a static black hole.
Indeed, if we assume that the black hole has a regular
horizon, then following the arguments in \cite{Bekenstein}
we arrive at the conclusion that the scalar field must be
constant. On the other hand, the inclusion of the scalar
field immediately results in singularities at the
horizon. Strictly speaking, there is no horizon anymore.
The surface of metric discontinuity becomes the surface
where the scalar curvature diverges. In this case the
arguments of \cite{Bekenstein} do not apply because the
surface integral diverges. This leaves the possibility for
the existence of the scalar field in the exterior of a black
hole if one admits singularities at finite radii.
However, on very general grounds it is unlikely that
such a surface of singularities can ever appear.

In this paper we will not discuss all these issues.
Here we simply assume that there is a scalar field
in the exterior of the electrically charged black hole and
study possible consequences of that. We will see that in
many cases the dilaton charge plays a role very similar
to that of mass. It will be shown that there is a
one-parameter family of solutions corresponding to
black holes with different dilaton charges. For a
particular choice of the dilaton charge the solution
reduces to the one obtained in \cite{GHS}, from now
on referred to as the GHS-solution. The general feature
of all these solutions is the presence of curvature
singularities at the horizons.

We will use geometrical units $c=G=1$ throughout the paper.
The line interval for a static spherically symmetric
space-time can be written as
\begin{equation}
ds^2=-\alpha^2 dt^2+\beta^2 dr^2+\gamma^2
(d\theta^2+\sin^2\!\theta\, d\varphi^2),
\label{interval}
\end{equation}
where $\alpha,\beta$ and $\gamma$ are functions of
the radial coordinate $r$ only. The solution that
corresponds to the black hole with mass $M$ and electric
charge $Q$ is given by the Reissner-Nordstr\"{o}m metric
with $\gamma^2(r)=r^2$ and
\[ \alpha^2(r)=\beta^{-2}(r)=\left( 1-{r_1\over r}\right)
\left( 1-{r_2\over r}\right) , \]
\[ {\rm where\ \ \ \ \ } r_{1,2}=M\pm \sqrt{M^2-Q^2}. \]
This solution has two horizons $r=r_{1,2}$
when $Q^2<M^2$. If $Q^2=M^2$ the two horizons coincide and
the black hole is said to be extremal. In the case when
$Q^2>M^2$ there are no horizons.

The metric is
divergent when $r$ approaches $r_1$ or $r_2$. However, invariant
quantities made out of components of the Riemann curvature tensor
are regular at the horizons. In particular, the scalar curvature
is zero everywhere except the origin, where the true
singularity resides.

The geometry of space-time is very different in the presence
of the dilaton. Even in the case of pure dilaton
gravity, when there is no
electromagnetic field, a singularity appears at a
finite radius.

Inclusion of the dilaton $\phi$ leads to the appearance of
a conserved charge, the
dilaton charge. In static space-time it is defined by
\[ D={1\over 4\pi}\oint\nabla_{\!k}\phi\,dS^k. \]
The integration is taken over a space-like surface enclosing
the origin. The conservation means that the value of the
dilaton charge does not depend on the
choice of the surface. This is a simple consequence of the
equation of motion for the scalar field:
$\nabla_{\!\mu}\nabla^{\mu}\phi=0$.

Static spherically symmetric solutions of the Einstein
equations and the dilaton equation are completely defined
by the dilaton charge $D$ and the mass $M$.
The metric components are given by
\begin{equation}
\alpha^2(r)=\beta^{-2}(r)=
\left( 1-{r_s\over r}\right) ^{2M\over r_s},
\label{pda}
\end{equation}
\begin{equation}
\gamma^2(r)=r^2\left( 1-
{r_s\over r}\right) ^{1-{2M\over r_s}}.
\label{pdg}
\end{equation}
The dilaton field is defined up to an arbitrary constant,
its value at infinity
\begin{equation}
\phi(r)=\phi_{\infty}+{D\over r_s}
\ln\left( 1-{r_s\over r}\right).
\label{pdf}
\end{equation}
The constant is irrelevant and can be set to zero.
The quantity
that plays the role of the Schwarzschild radius is defined by
\begin{equation}
r_s=2\sqrt{M^2+D^2}.
\label{pdh}
\end{equation}
Thus the metric shows a singularity at $r=r_s.$ Unlike the
Schwarzschild and the Reissner-Nordstr\"{o}m solutions this is
a true singularity. This can be seen from the formula for the
scalar curvature
\[ R\,(r)=\frac{2D^2}{r^{2+{2M\over r_s}}
(r-r_s)^{2-{2M\over r_s}}}. \]
Therefore, the scalar curvature becomes infinite as $r$
approaches $r_s$ for any nonzero value of the dilaton charge.

In the following sections we describe solutions that include
both a dilaton and an electromagnetic field. The solutions
share the main features of the Reissner-Nordstr\"{o}m
geometry and pure dilaton gravity. These are the
appearance of two horizons and the presence of a curvature
singularity at the horizons.

\section{Action and Symmetries}

The form of the  action in four dimensions is suggested
by the low-energy limit of string theory
\[ S=-{1\over 16\pi}\int {\sqrt{|g|}}\,
(R- 2\,\nabla_{\!\mu}\phi\,\nabla^{\mu}\phi -e^{-2\phi}
F_{\mu\nu}F^{\mu\nu})\,d^4x. \]
The equations of motion for the metric $g_{\mu\nu},$ the
vector potential $A_{\mu}$ and the dilaton field $\phi$ are
\begin{equation}
G_{\mu\nu} = 8\pi T_{\mu\nu},
\label{Einstein}
\end{equation}
\begin{equation}
\nabla_{\!\mu} (e^{-2\phi} F^{\mu\nu})=0,
\label{Maxwell}
\end{equation}
\begin{equation}
\nabla_{\!\mu}\nabla^{\mu}\phi+{1\over 2}e^{-2\phi}
F_{\mu\nu}F^{\mu\nu}=0.
\label{Dilaton}
\end{equation}
The components of the Einstein tensor and the energy-momentum
tensor are given in Appendix.

The symmetries of the action are general
covariance and gauge symmetry. In addition, the action
is invariant under the global scale transformations
\[ \tilde\phi(x)=\phi(x)+\Lambda, \]
\[ \tilde{A}_{\mu}(x)=e^{\Lambda} A_{\mu}(x). \]
This freedom can be eliminated by specifying
$\phi_{\infty}$, the value of the dilaton at infinity.
If nonzero this value will result in screening
of the electric charge
\[ E(r\rightarrow\infty)\simeq
{{Qe^{2\phi_{\infty}}}\over r^2}. \]
In what follows we assume that $\phi_{\infty}=0$.
The equations for
nonzero $\phi_{\infty}$ can be obtained by suitable
redefinitions.

The N\"{o}ther current corresponding to the global scale
transformations is
\[ J_{\mu}=\nabla_{\!\mu}\phi+e^{-2\phi}F_{\mu\nu}A^{\nu}. \]
Note that this current is not gauge invariant.
However the conserved charge, associated with the current,
is gauge invariant provided that only those gauge
transformations that vanish at infinity are allowed.

\section{Equations of Motion}

The metric for a static spherically symmetric space-time
is given by eq.~(\ref{interval}). This form of the
metric remains unchanged under
the following transformation, which
is a remnant of general coordinate invariance
\begin{equation}
r\rightarrow\tilde{r}{\rm\ \ \ \ and\ \ \ \ }\beta^2
\rightarrow\beta^2\left( {dr\over{d\tilde{r}}}\right) ^2.
\label{remnant}
\end{equation}
We will use this freedom to choose $\beta^2$ in such a way
that the Einstein equations of motion are simplified.

A spherically symmetric electric field is everywhere radial
\[ F_{rt} = f(r). \]
The Maxwell equations~(\ref{Maxwell}) can then be integrated
and give a generalization of the Gauss law for
curved space-time with the dilaton
\begin{equation}
e^{-2\phi}\gamma^2f=Q,
\label{Gauss}
\end{equation}
where $Q$ is electric charge.
The Einstein equations~(\ref{Einstein}) are not all
independent. This can easily be seen in the
orthonormal frame, where
$G_{\hat{\theta}\hat{\theta}}=
G_{\hat{\phi}\hat{\phi}}$.
The corresponding components of the energy-momentum
tensor also coincide (see Appendix).
We take the following linear combinations
of them as independent equations
\begin{equation}
G_{\hat{r}\hat{r}}+G_{\hat{\theta}\hat{\theta}}=0,
\label{E1}
\end{equation}
\begin{equation}
G_{\hat{t}\hat{t}}+G_{\hat{r}\hat{r}}=
2{\phi'^2\over \beta^2},
\label{E2}
\end{equation}
\begin{equation}
G_{\hat{t}\hat{t}}-G_{\hat{r}\hat{r}}=
2e^{-2\phi}\frac{f^2}{\alpha^2\beta^2}.
\label{E3}
\end{equation}
The prime denotes differentiation with respect to the
radial coordinate $r$. Note that eq.~(\ref{E1})
contains only metric fields. It can be written as
\begin{equation}
\left[ \frac{(\alpha\gamma)'\gamma}
{\beta}\right] '=\alpha\beta.
\label{EM1}
\end{equation}
This equation suggests the following gauge choice
\begin{equation}
\alpha\beta=1.
\label{C1}
\end{equation}
By making this choice we fix the freedom in eq.~(\ref{remnant}).
Then equation~(\ref{EM1}) can be integrated and gives
\begin{equation}
(\alpha\gamma)^2=(r-r_1)(r-r_2),
\label{C2}
\end{equation}
where $r_1$ and $r_2$ are arbitrary constants.
The situation when
both $r_1$ and $r_2$ are real and positive corresponds to
a black hole with two horizons located at $r=r_{1,2}$.
An extremal black hole arises when $r_1=r_2$. In the case
when both $r_1$ and $r_2$ are complex numbers, with $r_2$
the complex conjugate of $r_1$, the solution gives a naked
singularity. All this is in close analogy with the
Reissner-Nordstr\"{o}m solution.
We will often use the following linear
combinations of the parameters
\[ \bar{r}=\frac{r_1+r_2}{2} \ \ \ {\rm and}\ \ \
\Delta=r_1-r_2. \]
In the gauge, eq.~(\ref{C1}), the Einstein equations~(\ref{E2})
and (\ref{E3}) take the following form
\begin{equation}
\gamma''+\gamma\phi'^2=0,
\label{EM2}
\end{equation}
\begin{equation}
(\alpha^2\gamma\gamma')'=1-e^{-2\phi}\gamma^2f^2.
\label{EM3}
\end{equation}
The dilaton equation~(\ref{Dilaton}) also simplifies and
can be written as
\begin{equation}
(\alpha^2\gamma^2\phi')'=e^{-2\phi}\gamma^2f^2.
\label{D}
\end{equation}
The differential equations~(\ref{EM2})-(\ref{D})
together with the constraints~(\ref{C1}),(\ref{C2})
constitute a complete set of equations of motion.

To solve the equations of motion we reduce them to an
equation with only one field; this will be the dilaton
field. First combining equations~(\ref{EM3}) and
(\ref{D}), we obtain
\[ (\alpha^2\gamma^2\phi')'+
(\alpha^2\gamma\gamma')'=1. \]
Then integrating it and taking into account eq.~(\ref{C2})
we obtain
\begin{equation}
\phi'+{\gamma'\over \gamma} =
\frac{r-C}{(r-r_1)(r-r_2)}.
\label{Int}
\end{equation}
Here $C$ is an arbitrary constant of integration with
the dimensions of length. This parameter determines the type of
solution. By making the particular choice
$C=r_1({\rm or\ }r_2)$ one can remove the singularity
at $r=r_1({\rm or\ }r_2)$. However, it is not possible to
remove both singularities at the same time. We will
show that the constant $C$ is essentially given by the
dilaton charge of the solution.

Eliminating $\gamma$ between the
equations~(\ref{Int}) and (\ref{EM2}), we obtain
the equation with the dilaton field only
\begin{equation}
\phi''-2\,\phi'^2+
2\,\frac{r-C}{(r-r_1)(r-r_2)}\,\phi'=
\frac{(C-r_1)(C-r_2)}{(r-r_1)^2(r-r_2)^2}\, .
\label{main}
\end{equation}
This is an ordinary differential equation with singular
coefficients. The choice $C=r_1$ makes the
coefficient in front of $\phi'$
regular at $r=r_1$.
It also makes the right-hand side
of the equation vanish. In this case
the exact solution was found in \cite{GHS}.
A very similar solution may be found if one chooses
$C=r_2$. Here we allow $C$ to be arbitrary. With
arbitrary $C$ the dilaton equation~(\ref{main})
also can be integrated exactly.

Before we describe the integration let us introduce some
notation that will prove to be useful.
Instead of $C$ we will use $\sigma$ defined by the equation
\[ C=\bar{r}+\sigma. \]
It will be also convenient to introduce $\mu$ and
$\nu$ defined by
\[ \mu={\sigma\over \Delta}, \ \ \ \
\nu^2={1\over 2}-{\sigma^2\over \Delta^2}. \]
Note that $\mu$ and $\nu$ are not independent variables.

\section{Integration of the Dilaton Equation}

To integrate the dilaton equation~(\ref{main})
we first introduce a new radial coordinate
$\rho$ by the formula
\[ \rho=\frac{r-r_1}{r-r_2}. \]
Here we assume that $r_1$ and $r_2$ are real and
unequal with $r_1>r_2$.
Let us also introduce a new function $\psi(\rho)$
\[ \phi'(r)=\frac{(1-\rho)^2}
{2\rho\Delta}\,\psi(\rho). \]
These choices of
the new function and variable
greatly simplify the dilaton equation~(\ref{main}).
It can now be written as
\begin{equation}
\rho\,\frac{d\psi}{d\rho}=
(\psi+\mu)^2-\nu^2.
\label{simple}
\end{equation}
Then one can see that there are three different types
of solutions. They correspond to the following possibilities
\[ {\rm Type-I:}\ \ \ \ \ \ \nu^2>0, \]
\[ {\rm Type-II:} \ \ \ \ \ \nu^2<0, \]
\[ {\rm Type-III:}  \ \ \ \ \nu^2=0. \]
To complete the integration of the dilaton equation let us
write the integral of eq.~(\ref{simple}) for the Type-I solution
\begin{equation}
\psi+\mu-\nu=
\frac{2\nu\rho^{2\nu}}
{{\rm const}-\rho^{2\nu}}.
\label{Intpsi}
\end{equation}
All three types of solutions will be discussed in the
following sections.

\section{Solutions of Type-I}

The condition $\nu^2>0$ implies that
a solution of type-I has values of
the parameter $\mu$ that are restricted to the interval
$|\mu|<1/\sqrt2$.
In this range $\nu$
is a real number. We take $\nu$ positive. Integration of
eq.~(\ref{Intpsi}) gives
a solution for the dilaton field
\def\Vir1{r_1\rho^{-\nu}-r_2\rho^{\nu}}
\begin{equation}
e^{2\phi}=\frac{\rho^{-\mu}\Delta}{\Vir1}.
\label{d1}
\end{equation}
Knowing $\phi$, one can find a solution for the metric
components. For the first component $\alpha^2$ we obtain
\begin{equation}
\alpha^2=\frac{\rho^{\mu}\Delta}{\Vir1}.
\label{a1}
\end{equation}
Then the solution for the metric component $\gamma^2$ can
be found from eq.~(\ref{C2}). We rewrite this equation in
terms of the new radial coordinate
\begin{equation}
\alpha^2\gamma^2=\frac{\rho\Delta^2}{(1-\rho)^2}.
\label{g1}
\end{equation}
The solution for the electric field is a simple consequence
of the Gauss law, eq.~(\ref{Gauss})
\begin{equation}
f=\frac{(1-\rho)^2}{\rho}\frac{Q}{(\Vir1)^2}.
\label{f1}
\end{equation}
These equations are written for the exterior of the black
hole, $r>r_1$. General equations describing both the exterior
and the interior of the black hole would include absolute
values. In order not to complicate the equations, we assume
that $r>r_1$ and omit the absolute value signs whenever
possible.

The arbitrary constants $r_1$ and $r_2$, which correspond to
the horizons, can be found from the asymptotic behavior at
spatial infinity of the $g_{tt}$ component of the metric
\[ \alpha^2(r)\simeq 1-{2M\over r}. \]
One more condition is needed. The asymptotic behavior of the
electric field does not provide one, because it is satisfied
automatically. We take instead equation~(\ref{D}).
So the result is
\begin{equation}
r_1(\nu+\mu)+r_2(\nu-\mu)=2M,
\label{def1}
\end{equation}
\begin{equation}
2\nu^2 r_1 r_2 =Q^2.
\label{def2}
\end{equation}
These equations allow us, in principle, to find
$r_1$ and $r_2$.
However, one should be careful when using these
equations because $\mu$ and $\nu$
themselves depend on $r_1$ and $r_2$ through $\Delta$.
Apart from $M$ and $Q$, only $\sigma$ is a completely
arbitrary parameter.

\section{Dilaton Charge}

The solution for the pure dilaton black hole suggests
that the free parameter $\sigma$ must somehow be
related to the dilaton charge.
To calculate the dilaton charge one needs to know the
scalar potential $A_t$.
For the static electric field, eq.~(\ref{f1}), the
scalar potential is defined
up to an arbitrary constant, its value at infinity
\[ A_t(r)=A_t(\infty)-{Q\over{2\nu}}
\left( \frac{\rho^{-\nu}-\rho^{\nu}}
{\Vir1}\right). \]
Knowing the potential, one can find the components of
the dilaton current and calculate the flux
of the dilaton field through a closed space-like
surface. The result is
\[ {\rm Flux}=\oint J_k\, dS^k=
4\pi (M-\sigma-QA_t(\infty)). \]
Note that the flux does not depend on the choice of the surface.
The dilaton charge is defined
as the flux per unit solid angle.
Assuming that the electric potential vanishes at infinity, we
obtain the relation between the free parameter $\sigma$ and
the dilaton charge
\[ D=M-\sigma. \]
Thus we see that the arbitrary constant $C$ is linearly
related to the dilaton charge.
The formula for the dilaton charge, although quite simple,
has many consequences. One of them is the possibility of
having a black-hole solution with two horizons even when
the mass is zero.

\section{The Horizons and the Barrier}

Let us analyze the horizons of the black hole with
dilaton charge. To find the locations of the
horizons we need to solve equations~(\ref{def1})
and (\ref{def2}), which we rewrite now as
\begin{equation}
\nu(r_1+r_2)=M+D,
\label{defm1}
\end{equation}
\begin{equation}
2\nu^2 r_1 r_2=Q^2.
\label{defm2}
\end{equation}
Note that $\nu$ depends on $r_1$ and $r_2$
through $\Delta$.
One way to solve these equations is
to find $\Delta$ first. Simple calculations lead to
\begin{equation}
\Delta^2=4(M^2+D^2-Q^2).
\label{delta}
\end{equation}
Knowing $\Delta$, we can now solve
equations~(\ref{defm1}), (\ref{defm2}) for
$r_1$ and $r_2$. The result is
\begin{equation}
r_{1,2}=\sqrt{M^2+D^2-Q^2}\left[
\pm 1+\frac{M+D}{\sqrt{(M+D)^2-2Q^2}}\right].
\label{horizons}
\end{equation}
The sign ``+" corresponds to $r_1$.
We see that
both $r_{1,2}$ are real and positive provided that
\[ Q^2<{1\over 2}(M+D)^2. \]
This is the region of validity of the type-I
solution we described earlier. Thus we see that
the space-time around
the type-I black hole has two horizons.

If we increase the electric charge so that it approaches
the critical value given by
\begin{equation}
Q^2={1\over 2}(M+D)^2,
\label{critical}
\end{equation}
both horizons extend to infinity and $\nu^2\rightarrow+0$.
Therefore in the limit this particular case
corresponds to the solution of type-III.

If we further increase the electric charge so that
\begin{equation}
{1\over 2}(M+D)^2<Q^2<M^2+D^2,
\label{boundaries}
\end{equation}
we find that both $r_{1,2}$ become complex numbers with
the same imaginary part. That means $r_1$ and $r_2$ are
not complex conjugate to each other.
Now one can see that $\nu^2<0$, and we are
in the region of the type-II solution.

Finally, the electric charge can be so large that
\[ Q^2>M^2+D^2. \]
Here again $r_{1,2}$ are complex numbers, but now
$r_2$ is the complex conjugate of $r_1$. Of course,
$\Delta$ becomes pure imaginary. As a consequence
of this $\nu^2>0$ and we again have a
solution of type-I, but now without
horizons.

An important observation comes from the condition in
eq.~(\ref{boundaries}). Namely, the upper limit
is always greater than or equal to
the lower limit. This leads to the existence of a
barrier that separates the two regions of the type-I
solution. In either of these regions the solution with
given mass and electric charge is
a one-parameter family with arbitrary values of
the dilaton charge. We will see that
inside the barrier there are no solutions unless
the dilaton charge takes a special value.

When the upper limit coincides with
the lower limit, the solution of type-I with two
horizons can be continuously transformed into the solution
of type-I with no horizons. This is only possible when
the barrier shrinks to a single point. This happens
if the dilaton charge takes the value
\[ D=M. \]
For all other values of the dilaton charge there is a
finite barrier, the region where real solutions are
forbidden.

The different regions and the boundaries between them
are shown in Figure. It is convenient to take
the pair $(D,Q^2)$ as cartesian coordinates on the
plane. Then the curves corresponding to
the boundaries in eq.~(\ref{boundaries}) are
two parabolas, one being above the other. The barrier
is the region in between the curves (the shaded area
in Figure). Above and below the barrier there are
solutions of type-I. The barrier shrinks
to a point when the two curves touch each other.
This is the point with coordinates $(M,2M^2)$. At this
point the curves have a common tangent line defined by
\[ Q^2=2MD. \]
Inside the barrier real solutions do not exist, except
when $D=-M$. This segment of a vertical line corresponds
to the solution of type-II.

\section{Duality of Type-I Solutions}

There is one feature of a solution of type-I which is
peculiar to dilaton black holes only. This is a
symmetry between the mass and the dilaton charge of the
solution. Let us take a solution of type-I which is
characterized by mass $M$ and dilaton charge $D$.
If we interchange $M$ and $D$, we obtain another solution
of type-I. The equations for the metric and dilaton fields
of this solution can be obtained from the corresponding
equations of the former solution by the substitution
$\mu\rightarrow-\mu$. The new solution is quite different
from the old one, though both solutions have horizons at
the same locations. The two solutions are dual to
each other in the sense that the role of the mass in one
solution is played by the dilaton charge in the other.
The relation between the two solutions leads us
to think of them as parts of one solution. Namely, suppose
the solution of the type-I above has $D<M$. By gradually
increasing the dilaton charge to values $D>M$, we
continuously transform the solution into its dual. Of
course, at the moment when $D=M$ the solution and its dual
coincide. Examples of dual solutions are given in
the next section.

\section{Special Cases of Type-I Solutions}

It is interesting to consider various special cases of
type-I solutions with two horizons. First let us set
one of the parameters to zero, while the other two remain
unrestricted.

\noindent
(i) $Q=0$.

\noindent
When the electric charge vanishes the inner horizon
shrinks to a point. For the outer horizon we find that
$r_1=r_s$ (see eq.(\ref{pdh})), and the solution reduces
to the one of pure dilaton gravity,
eqs.~(\ref{pda})-(\ref{pdf}). This is not surprising
because setting $Q=0$ is equivalent to dropping the
Maxwell term in the action.

\noindent
(ii) $D=0$.

\noindent
When the dilaton charge vanishes, the solution is
specified by mass and electric charge, as in the
case of the Reissner-Nordstr\"{o}m black hole. However,
it is very different from the latter. One of the
differences is that values of the electric
charge in the interval $M^2\!/2<Q^2<M^2$ are
not allowed. We will also see that the curvature becomes
singular at the horizons.

\noindent
(iii) $M=0$.

\noindent
Unlike other classical black holes, the dilaton black
hole can exist without mass. The role of the mass in
the formation of the horizons is now played by the
dilaton charge. This solution is dual to the previous
one.

\noindent
(iv) $Q^2=2MD$

\noindent
Let us now consider a special case of the solution
of type-I that corresponds to the tangent line.
In this case the horizons are given by particularly
simple expressions. For $D<M$ they are
\begin{equation}
r_1=2M\ \ \ \ {\rm and}\ \ \ \ r_2=2D.
\label{ghshor}
\end{equation}
The metric also becomes very simple
\begin{equation}
ds^2=-\left( 1-{2M\over r}\right) dt^2+
\left( 1-{2M\over r}\right)^{-1} dr^2+
r^2\left( 1-{2D\over r}\right) (d\theta^2+
\sin^2\!\theta\,d\varphi^2).
\label{ghsint}
\end{equation}
The dilaton is formally independent of mass
\begin{equation}
\phi(r)={1\over2}\ln\left( 1-{2D\over r}\right).
\label{ghsdil}
\end{equation}
This is the GHS-solution. It has the remarkable property
of being regular at the outer horizon. As we noted earlier,
this solution corresponds to the choice $C=r_1$ in the
dilaton equation.

Quite naturally, for $D>M$ we find that the horizons  are
switched
\begin{equation}
r_1=2D\ \ \ \ {\rm and}\ \ \ \ r_2=2M,
\label{dualhor}
\end{equation}
while the metric and the dilaton are given by the same
expressions as before. This is the dual solution. It
corresponds to the choice $C=r_2$, and it is regular at
the inner horizon.

\section{Curvature Singularities}

As we already saw in the case of pure dilaton
gravity, the scalar curvature is divergent at a finite
value of the radial coordinate. This is
also true for the general solution of type-I. One way
to study the behavior of the curvature is to use the
equations of motion. Note that when the equations of
motion are satisfied we have a simple formula for the
scalar curvature, $R=2\alpha^2\phi'^2$ (see Appendix).
Using this formula, one can see that when
$r\rightarrow r_1\!+\!0$,
or equivalently $\rho\rightarrow+0$,
the curvature diverges as
\[ R\propto (\mu-\nu)^2 \rho^{-n}, \]
\[ {\rm where\ \ }n=2-\mu-\nu\geq 1. \]
At the other horizon $r\rightarrow r_2\!-\!0$,
or equivalently
$\rho\rightarrow+\infty$,
the scalar curvature is also divergent
\[ R\propto (\mu+\nu)^2 \rho^m, \]
\[ {\rm where\ \ }m=2+\mu-\nu\geq 1. \]
These equations show that generically the
scalar curvature is divergent at both horizons.
One can make the curvature regular at
$r=r_1$ by setting $\mu=\nu$. This condition leads
to the GHS-solution. Therefore the only
solution of type-I that has regular curvature
at $r=r_1$ is the GHS-solution. Similarly one can set
$\mu=-\nu$ and obtain the solution with regular
curvature at $r=r_2$. This is the dual solution.
It is also clear that there is no
solution for which the scalar curvature is regular
at both horizons.

\section{Solutions of Type-II and Type-III}

We know that when $Q^2$ takes the critical value
given by eq.~(\ref{critical}) the parameter $\nu$
vanishes and we should obtain a solution of type-III.
In principle, we should be able to find the equations
for the metric components and the dilaton field
of the type-III solution from
the corresponding equations of the type-I solution
by taking the limit $\nu^2\rightarrow+0$.
However as $Q^2$ approaches the critical
value both horizons extend to infinity while
$\Delta$ remains finite. The immediate consequence
is that in the limit $\rho=1$ for all values of $r$. One can
see then that some equations, for example eq.~(\ref{g1}),
become undefined. Thus the limit $\nu^2=0$ is divergent and
there are no solutions of type-III.

Now let us turn to the solution of type-II, i.e.
$\nu^2<0$. This solution appears only when the
condition in eq.~(\ref{boundaries}) is satisfied.
This is the region inside the barrier.
Here $r_1$ and $r_2$ are complex but not complex
conjugate to each other. Rather, they are related by
$r_2=-r_1^*$. This means that the sum $r_1+r_2$ is
pure imaginary. This situation is very undesirable,
because the quadratic form in
eq.~(\ref{C2}) must be real. Remember that the quadratic
form is a product of the metric components
$(-g_{tt})g_{\theta\theta}$.
Therefore the only possibility for a real solution
to exist in this region is for $r_2=-r_1$. This
can happen if the dilaton charge takes the special
value
\[ D=-M. \]
Knowing this, it is easy to construct the solution
of type-II from the known solution of type-I.
The result is
\def\Vir2{\cos(\lambda\ln\rho)}
\[ e^{2\phi}=\frac{\rho^{-\mu}}{\Vir2}, \ \ \
\alpha^2=\frac{\rho^{\mu}}{\Vir2}, \]
where $\lambda$ is a real root of $\lambda^2=-\nu^2$.
Again, $\gamma^2$ can be found from eq.~(\ref{g1}).
The solution above is written in terms of the radial coordinate
$\rho$ now defined by
\[ \rho=\frac{r-r_1}{r+r_1}. \]
The black hole described by this solution has only one
horizon for all allowed values of the electric charge. It is
located at
\[ r_1=\sqrt{2M^2-Q^2}. \]
Thus we see that inside the barrier the solution
exists only if the dilaton charge ``compensates'' the
mass. For all other values of the dilaton charge
there are no real solutions in this region.

\section{Solution of Type-I Above the Barrier}

The last possibility to consider is
the region above the barrier:
\[ Q^2>M^2+D^2. \]
In this case both $r_1$ and $r_2$ are complex
and one is complex conjugate to the other.
Let us denote $r_1=u+iv$, where $u$ and $v$ can
be found from eq.~(\ref{horizons}).
Then the radial coordinate
$\rho$ becomes a phase on the complex plane
\[ \rho=\frac{r-r_1}{r-r_1^*}=e^{-2i\eta}, \]
\[ {\rm where}\ \ \ \  \tan\eta={v\over{r-u}}.\]
Knowing this, one can readily obtain a solution
for the dilaton field
\def\Vir3{\cos2\nu\eta+{u\over v}\sin2\nu\eta}
\[ e^{2\phi}=\frac{e^{{\sigma\over v}\eta}}{\Vir3}, \]
and the first metric component
\[ \alpha^2=\frac{e^{-{\sigma\over v}\eta}}{\Vir3}. \]
The other metric component $\gamma^2$, again, can be
found from eq.~(\ref{C2}). This equation written in terms
of $\eta$ takes the following form
\[ \alpha^2\gamma^2=\frac{v^2}{\sin^2\!\eta}. \]
The solution has no horizon at all.
Since $\nu^2>0$, it is type-I, though quite different
from the type-I solution we described above.

\section{Extremal Black Holes}

In Sect.7 we obtained the expression for $\Delta$ in
terms of the mass and the electric and dilaton charges.
Extremal black holes appear when $\Delta$ vanishes. One can
see that, no matter where the horizon is located, extremality
occurs when
\begin{equation}
Q^2=M^2+D^2.
\label{extremal}
\end{equation}
If we approach extremality from above
with some arbitrary value of the dilaton charge we
inevitably obtain that $r_1=r_2=0$. If we try to
approach extremality from below by increasing the
electric charge in the  solution of type-I,
then we encounter the barrier well before we
meet extremality (see Figure).
By approaching extremality
along the line of the solution of type-II we again
obtain that $r_1=r_2=0$.
The only possibility that remains is to approach the
extremal curve, eq.~(\ref{extremal}), at
the point where the barrier disappears. This is the
only point where the solution of type-I with two
horizons can reach extremality.
Thus we see that an extremal black hole can have a
nonvanishing horizon only if the extremality
condition eq.~(\ref{extremal}) is supplemented by
\[ D=M. \]
As we know, the point is located at $(M,2M^2)$.
If we approach this point from anywhere below the
tangent line, $Q^2=2MD$, we obtain the solution
\begin{equation}
\alpha^2=e^{2\phi}=\frac{r-r_1}{r+r_0},
\label{ext1}
\end{equation}
\begin{equation}
\gamma^2=(r+r_0)(r-r_1).
\label{ext2}
\end{equation}
Here the horizon $r_1$ and the shift $r_0$ are given by
\[ r_1=\sqrt2 M\ \ \ {\rm and}\ \ \  r_0=(2-\sqrt2)M. \]
We can also approach the point by moving along the
tangent line, see eqs.~(\ref{ghshor})-(\ref{dualhor}).
This way we obtain the same solution, namely
eqs.~(\ref{ext1})-(\ref{ext2}), but with zero shift
\[ r_1=2M\ \ \ {\rm and}\ \ \ r_0=0. \]
It is clear that we cannot redefine the radial
coordinate in the solution of type-I so that
both procedures would lead to the solution with zero shift.
The ambiguity in the shift comes from the fact that the orders
of singularity of the coefficients in the dilaton
equation~(\ref{main}) change abruptly when $r_1=r_2$.
The value of the shift is unphysical due to the symmetry
$\ \tilde{r}=r+{\rm const}\ $ and can be set to zero.
Thus we see that at the point $(M,2M^2)$ the extremal solution
indeed has a finite horizon. The horizon, however, is singular.
Since the scalar curvature diverges at the horizon (see Sect. 10)
it is a singularity rather than an event horizon.

\section{Arbitrary Coupling Constant}

To study the dependence of the dilaton black hole
solution on the strength of interaction between the
dilaton and the electromagnetic field, an arbitrary
coupling constant $a$ was introduced in \cite{GHS}.
This parameter modifies the Maxwell term in
the action in the following way
\[ e^{-2a\phi}F_{\mu\nu}F^{\mu\nu}. \]
The introduction of this coupling constant makes
it possible to have both weak $(a<<1)$ and strong
$(a>>1)$ coupling regimes. Then the solution should
reduce to the Reissner-Nordstr\"{o}m
solution in the limit $a=0$. The choice $a=1$
emerges from string theory. We already described
all the possible black hole solutions in the case
$a=1$. The overall picture
described above changes very little when the constant
$a$ is allowed to take arbitrary values.
Again there is a finite barrier that separates two
regions of the solution of type-I. The boundaries
of the barrier are given by
\[ \frac{(M+aD)^2}{a^2+1}<Q^2<M^2+D^2. \]
Inside the barrier a solution exists only if
the dilaton charge takes the special value $D=-M/a$.
The solution in this case will be type-II.
The boundaries of the barrier touch each other
at the point
\[ D=aM\ \ \ {\rm and}\ \ \ Q^2=(a^2+1)M^2. \]
Again, this is the only point where an extremal
solution can have a nonvanishing horizon.

Explicit formulas show that the solution of
type-I indeed reduces to the Reissner-Nordstr\"{o}m
solution in the limit $a=0$.
The solution of type-II exists only for $a>0$.

\section{Concluding Remarks}

We have seen that the dilaton field drastically affects
the space-time geometry of electrically charged black
holes. The inclusion of the dilaton almost inevitably
destroys the horizons of the Reissner-Nordstr\"{o}m solution.
The only exception is the GHS solution, which has a regular
outer horizon but whose inner horizon is singular.
In this paper we described all possible solutions assuming
that the dilaton charge can take arbitrary values. It is
interesting that we found the barrier, the region where no
real solution can exist. We also have seen that the dilaton
charge plays a role quite similar to that of the mass.
Moreover, we found that the simple interchange of the mass and
the dilaton charge in one of the solutions gives us another
solution of the same set of equations of motion. This fact we
called duality. These are all effects of the dilaton field.
Although the existence of the dilaton black holes is
problematic due to the curvature singularities, the exact
solutions we described above provide a simple framework for
studying different effects of the dilaton in general relativity.

\section*{Acknowledgments}

The author wishes to thank Prof. John Schwarz for
interesting conversations and comments on this paper.

\newpage

\section*{Appendix}

It is convenient to write the components of various tensors
in the orthonormal frame defined by the tetrad
\[ e^{\hat\mu}_{\nu}={\rm diag}\{\alpha,\,\beta,\,\gamma,\,
\gamma\sin\theta\}. \]
In this frame the nonzero components of the Riemann curvature
tensor for the metric of eq.~(\ref{interval}) are
\[ R_{\,\hat{t}\,\hat{r}\,\hat{t}\,\hat{r}}=
{1\over {\alpha\beta}}\left( {\alpha'\over\beta}\right)', \]
\[ R_{\,\hat{t}\,\hat{\theta}\,\hat{t}\,\hat{\theta}}=
R_{\,\hat{t}\,\hat{\phi}\,\hat{t}\,\hat{\phi}}=
{1\over\beta^2}\frac{\alpha'\gamma'}{\alpha\gamma}, \]
\[ R_{\,\hat{r}\,\hat{\theta}\,\hat{r}\,\hat{\theta}}=
R_{\,\hat{r}\,\hat{\phi}\,\hat{r}\,\hat{\phi}}=
-{1\over{\beta\gamma}}\left( {\gamma'\over\beta}\right)', \]
\[ R_{\,\hat{\theta}\,\hat{\phi}\,\hat{\theta}\,\hat{\phi}}=
{1\over\gamma^2}-\frac{\gamma'^2}{\beta^2\gamma^2}. \]
The components of the Ricci tensor and the scalar curvature
can be found from these formulas by contraction.
The Einstein tensor has the following nonzero components
\[ G_{\hat{t}\hat{t}}={2\over\beta^2}\left(
\frac{\beta'\gamma'}{\beta\gamma}-{\gamma''\over\gamma}
\right) +{1\over \gamma^2}-
\frac{\gamma'^2}{\beta^2\gamma^2}, \]
\[ G_{\hat{r}\hat{r}}=
{2\over\beta^2}\frac{\alpha'\gamma'}{\alpha\gamma}
-{1\over\gamma^2}+\frac{\gamma'^2}{\beta^2\gamma^2}, \]
\[ G_{\hat{\theta}\hat{\theta}}=
G_{\hat{\phi}\hat{\phi}}=
{1\over\beta^2}\left( {\alpha''\over\alpha}-
{{\alpha'\beta'}\over{\alpha\beta}}+
{{\alpha'\gamma'}\over{\alpha\gamma}}-
{{\beta'\gamma'}\over{\beta\gamma}}+
{\gamma''\over\gamma}\right). \]
In the orthonormal frame the components of the
energy-momentum tensor are especially simple
\[ 8\pi T_{\hat{\mu}\hat{\nu}}={\phi'^2\over\beta^2}
\left( \begin{array}{cccc}
1 & \  & \  & \  \\
\  & 1 & \  & \  \\
\  & \  & -1 & \  \\
\  & \  & \  & -1 \end{array} \right) +
e^{-2\phi}{f^2\over{\alpha^2\beta^2}}
\left( \begin{array}{cccc}
1 & \  & \  & \  \\
\  & -1 & \  & \  \\
\  & \  & 1 & \  \\
\  & \  & \  & 1 \end{array} \right). \]
The electromagnetic part of the energy-momentum tensor
is traceless. Thus the only contribution to the scalar
curvature comes from the dilaton field. By taking the
trace of the Einstein equation one finds that
$R=2\beta^{-2}\phi'^2.$
This equation provides a simple way to obtain
the scalar curvature for a metric that is a solution
of Einstein equations.

\newpage

\epsfbox{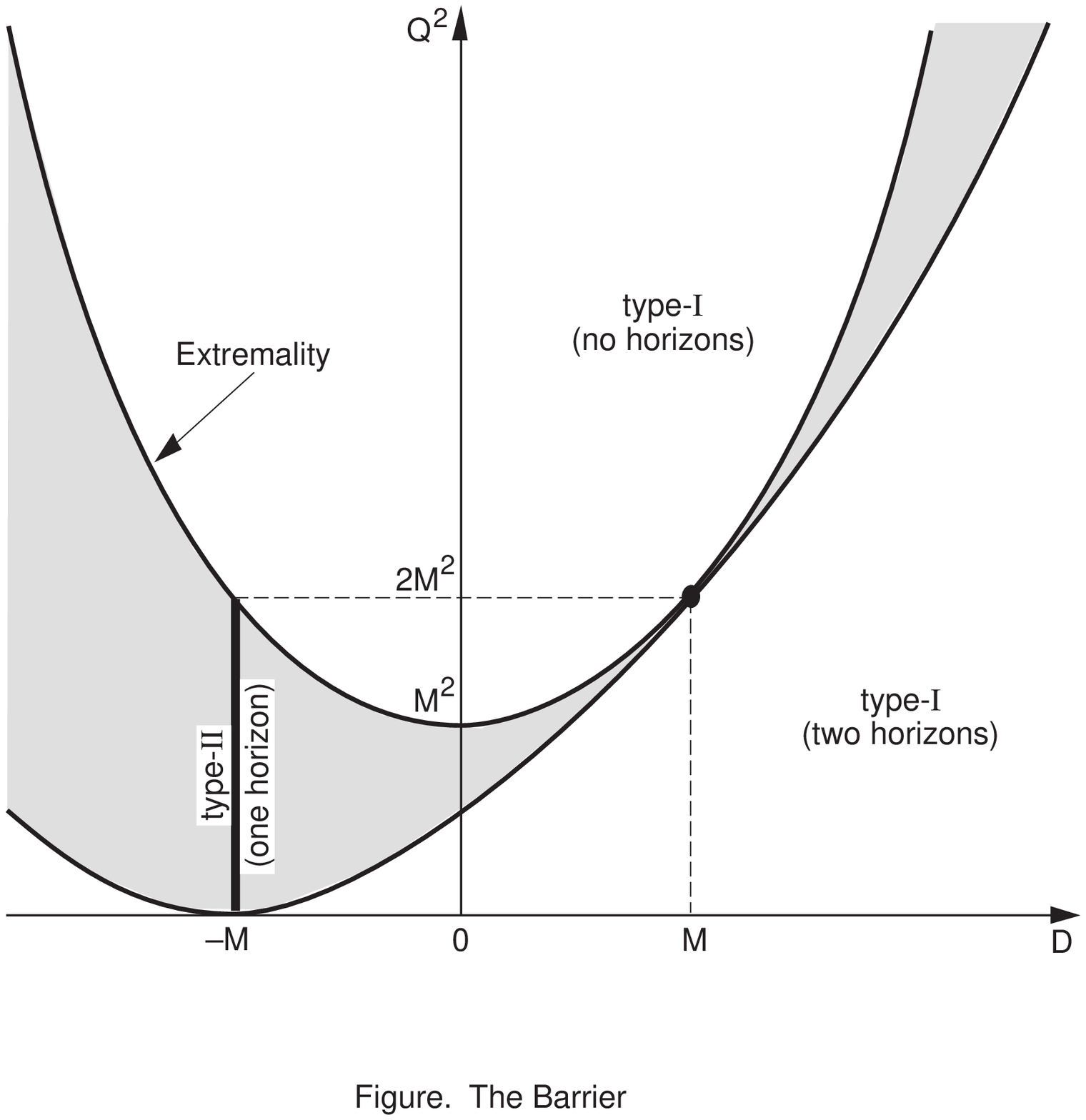}

\end{document}